\documentclass[graybox]{svmult}

\usepackage{mathptmx}       
\usepackage{helvet}         
\usepackage{courier}        
\usepackage{type1cm}        
\usepackage{makeidx}         
\usepackage{graphicx}        
\usepackage{multicol}        
\usepackage[bottom]{footmisc}

\makeindex             

\begin{document}

\title*{Light pollution in Spain. An european perspective}
\author{Alejandro S\'anchez de Miguel and Jaime Zamorano}
\institute{ \at Universidad Complutense de Madrid \email{alex,jaz@astrax.fis.ucm.es}}

\maketitle

\abstract{Spain appears in light pollution maps as a country less polluted than
their neighbours in the European Union. This seems to be an illusion
due to its low population density. The data indicate that Spain is one
of the most contaminated countries. To reach these conclusions we
compare the Spanish case to those of other European countries.}

\section{European light emission to the space}
\label{sec:1}

Using DMPS satellite night images taken in year 2000 we have estimated the
saturated surface in many European countries. The goal of this study
is to compare the illumination conditions and its effects in light
pollution. 

We have measured the surface of the saturated zone. We here compared
this parameter to the ratio of population density of the country. To
avoid biases we normalize by dividing the area built in each country
prior to the comparison. In this way it only takes into account the
surface that really makes sense to illuminate.

\begin{figure}[t]
\includegraphics[width=1.\textwidth,viewport=70 80 770 510,clip]{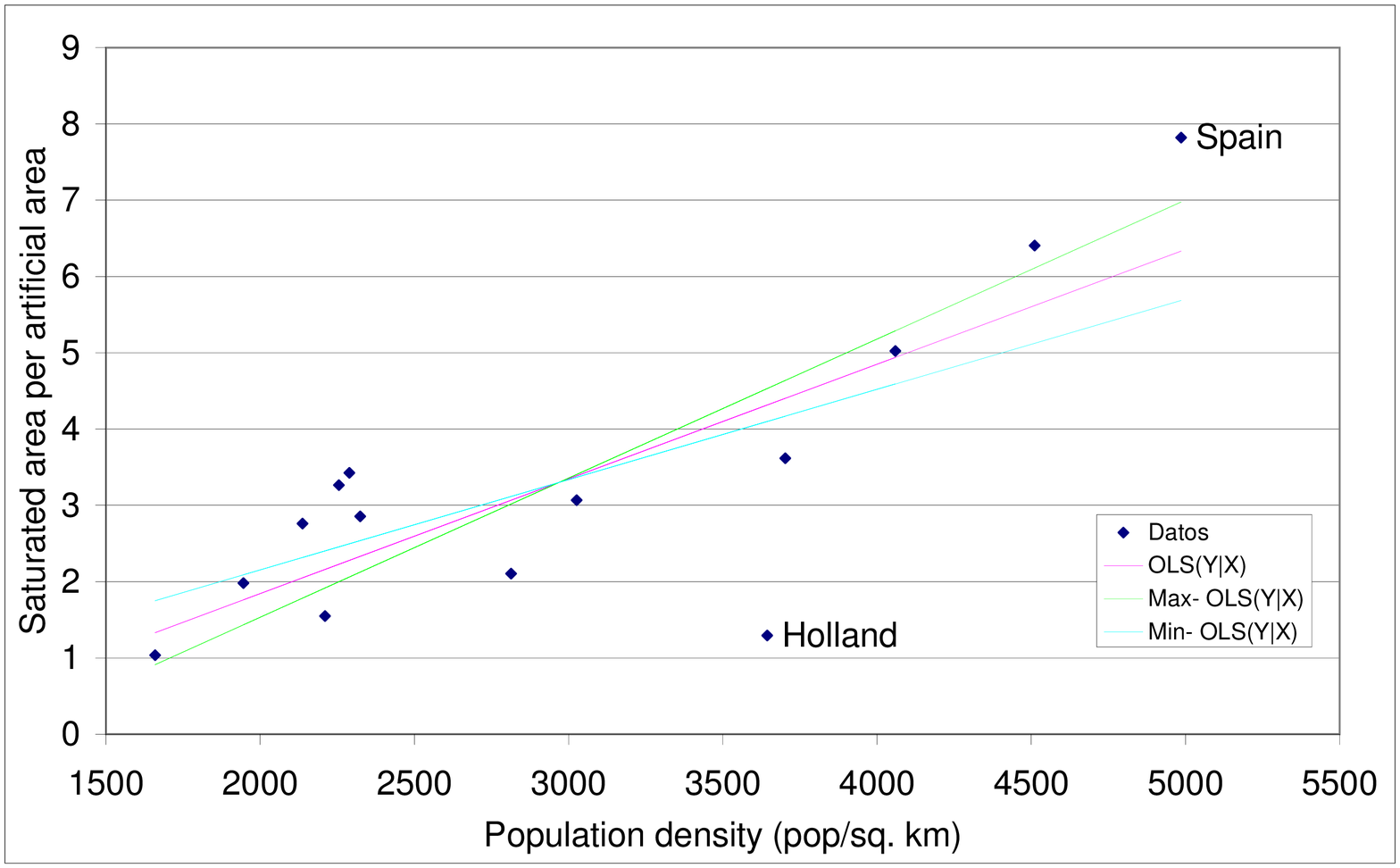}
\caption{The ratio of saturated surface area over
  built area versus the ratio of population density per built area. Data have been taken from  
  European Enviroment Agency EEA (www.eea.europa.eu).}
\label{fig:1}    
\end{figure}

\begin{figure}[t]
\includegraphics[width=1\textwidth, viewport=70 80 770 490,clip]{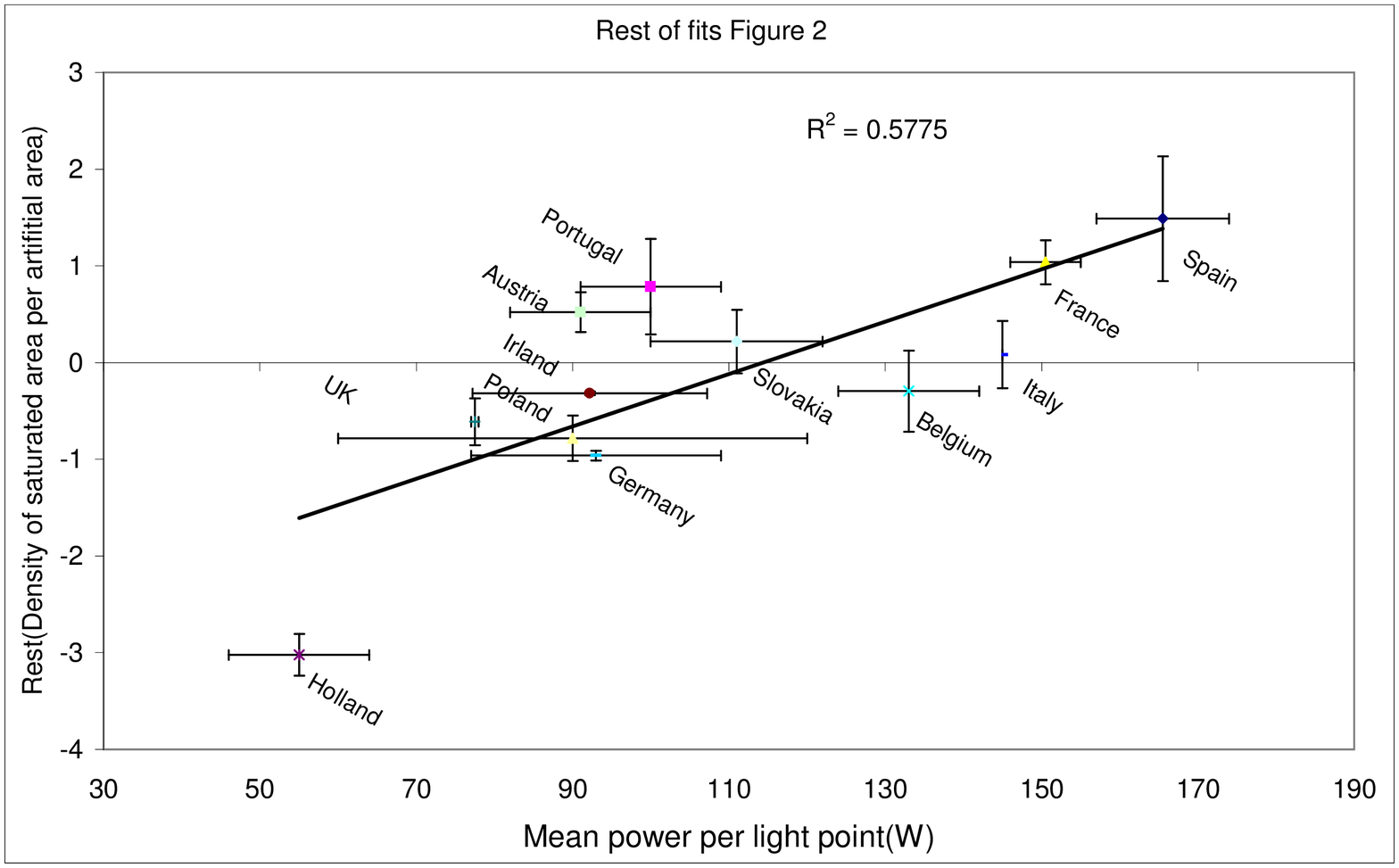}
\caption{Residuals of the Fig. 1 fits. Its shows a
relationship exists between the excess in Fig. 1b and the mayor power
installed in Spain. This relationship is statistically
significant. Correcting this effect, the dispersion on figure 2 goes
from R$_2$ 0.56 to 0.81.}
\label{fig:1}    
\end{figure}

In satellite images and other light pollution (LP) maps, Spain seems
to have less light pollution than other European countries. That could
be an effect caused by the low population density. We have looked for
the best intensive demographic parameter that represents better the
real LP.  This study shows that the correction using the mean light
power used is important as non demographic parameter to take into
account.

The results differ from the preliminary study of S\'anchez 2007
because now we use high resolution images and new realised data of
built area. Fits proposed by Isobe et al. (1990) have been
used. Fig. 1 shows the ordinary least squares fit and its outstanding
high and low for the uncertainty of this magnitude. The partial
correlation corrected by power installed by light point, increase the
correlation from R$_2$=0.56 to R$_2$=0.81. It can be discarded not
correlation with 95\% of confidence using the tests of Kendal, Fisher
and Spearman. Bootstrap simulations and no parametric tests have also
been used. In Fig. 1, Holland and Spain have been marked because are
the extreme examples.

\begin{figure}[t]
\hspace{-1cm}
\includegraphics[width=1.15\textwidth]{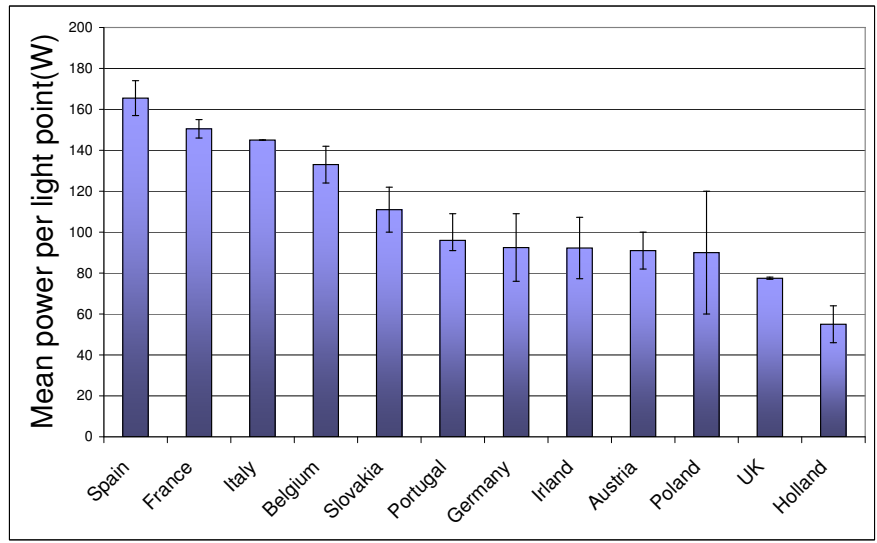}
\vspace{-1cm}
\caption{Mean power used per light point in EU countries. These data
  have been taken from two studies of the EU (Van Tichelen et al 2007;
  EnLight Project). In the Spanish case, the data are from Instituto
  de la Diversificaci\'on de la Energ\'{\i}a (IDEA) and Lamps
  Manufacturers Association (CLEMA).}
\label{fig:3}    
\end{figure}

\section{Spain: Champion of Europe in power for lights}

\subsection{Excess saturated area in Spain}

Spain and Holland position in Fig. 1 show a very important excess if
it is compared to its theoretical position according with the fits.  It is
possible that other sources of error not connected demographics
influence cause this dispersion. But we have found a correlation
between power installed by light point and residuals of the fits between
saturated and built area (see Fig. 2).

Now we are looking for other factors that could be connected as Light
point's density.  A possible answer: If we consider the same sort of
lamp, an increased power always produce greater emission.  While the
type use of the lamp and lighting can affect the final flux, the
energy available is always marked by the power of the lamp. The data
of EU studies (Van Tichelen et al 2007; EnLight Project) show that
Spain is the country that used the most powerful lamps So, we have
good reasons to suppose that the power excess in the Spanish cities
makes that in Spain around 1,5 km$^2$ more saturated area per square
kilometre built.

The power excess increasing the effect of an incorrect use of light
points and the light flux reflected of the pavements. Without control
of power installed it will be impossible to reduce the increase effect
of LP.  The actual LP models remark the effect important secondary
diffusion (Aub\'e 2007). It makes that great power light concentrated
are worst than the same in disperse area.

\subsection{Growing concern}
The Energy General Secretary of Spain data show that the growing of
installed power is constant since 1967 (Publicationes of the Spanish
Ministerio de Industria, Turismo y Comercio, see references).
Apparently, there is a plateau during 8 years, but detailed analysis
show important increases in Madrid, Barcelona and other provinces
consumption being the flat zone an artifact caused by the change of
the statistical processes. The bottom line is that, despite the fight
against Light Pollution, we are losing the battle.

{}

\begin{acknowledgement}
Thanks to J. Gorgas and N. Cardiel with the statistical treatment.
\end{acknowledgement}

\end{document}